\documentclass[aps,showpacs]{revtex4}
\usepackage{bm}
\usepackage{graphicx}
\usepackage{subfigure}
\begin{document}
 
\title{Sub-Kolmogorov-Scale Fluctuations in Fluid Turbulence}
\author{J\"org Schumacher}
\affiliation{Department of Mechanical Engineering,
           Technische Universit\"at Ilmenau, D-98684 Ilmenau, Germany}

\begin{abstract}
We relate the intermittent fluctuations of velocity gradients in turbulence
to a whole range of local dissipation scales generalizing the picture of a single mean
dissipation length. The statistical distribution of these local dissipation scales as a 
function of Reynolds number is determined in numerical simulations of forced homogeneous 
isotropic turbulence with a spectral resolution never applied before which exceeds the 
standard one by at 
least a factor of eight. The core of the scale distribution agrees well with a theoretical
prediction. Increasing Reynolds number causes the generation of ever finer local dissipation
scales. This is in line with a less steep decay of the large-wavenumber energy spectra 
in the dissipation range. The energy spectrum for the highest accessible Taylor microscale 
Reynolds number $R_{\lambda}=107$ does not show a bottleneck.
\end{abstract}

\date{\today}

\maketitle

\section{Introduction}
Turbulence is characterized by large fluctuations of velocity gradients
which appear preferentially at the smallest scales of the flow. The amplitudes
of these events exceed the mean values by orders of magnitude which is known
as small-scale intermittency \cite{Kolmogorov1962,Sreenivasan1997}. It is also
believed that these intensive fine-scale fluctuations are intimately connected
with the nonlinear cascade-like transfer of kinetic energy through the hierarchy
of eddy structures that fill the fluid on larger scales
\cite{Paladin1987,Nelkin1990,Frisch1991,Chevillard2005,Schumacher2007}.
A better understanding of fluid turbulence as a whole requires thus a detailed
resolution of the intermittent dynamics at the small-scale end of the inertial range.
In other words, it is necessary to determine how deep into the beginning
dissipation range the roughest filaments from the inertial 
range can sweep and how they affect the spectral decay of turbulent fluctuations with 
increasing Reynolds number. This can also be important for 
the mixing in reacting and non-reacting flows at high Schmidt number 
\cite{Sreenivasan2004,Schumacher2006} where a significant fraction 
of stirring of the concentration field takes place in the dissipation range 
of the flow. Such a study requires that the steepest gradients and their 
statistics are well resolved. Although significant progress in measurement 
techniques has been made \cite{Lathrop2003}, the finest structures remain still 
spatially unresolved in experiments. 

In this Letter, we present numerical simulations of forced homogeneous isotropic turbulence 
that unravel the dynamics in exactly the region where inertial and dissipation ranges match,
both in physical and Fourier space. 
In order to study the fine-scale structure and their statistics, a grid resolution is applied
which exceeds that of standard simulations by an order of magnitude. In particular, the following 
questions will be answered. What are the finest spatial scales across which large-amplitude 
gradient events evolve? How are these scales distributed as a function of the Reynolds number? 
Is consequently the large-wavenumber behaviour of the energy spectra in the dissipation range 
Reynolds number-dependent? 

The large-wavenumber behaviour 
of the turbulent fluctuations at the smallest scales is a long-standing 
problem. Kolmogorov postulated 65 years ago a universal form of the decay of the
energy spectrum that goes 
deep down into the dissipation range \cite{Kolmogorov1941}. Since then several 
analytical attempts have been made to determine the spectrum. The works left 
unspecified constants \cite{Heisenberg1948,Kraichnan1959,Foias1990,Gagne1991} or 
considered an infinitely extended range of excited scales \cite{Sirovich1994,Lohse1995}. 
Numerically, the time advancement in the dissipation range is very challenging 
since a significant fraction of the computational ressources has to be spent for the resolution of 
the tiny amplitudes \cite{Chen1993,Martinez1997,Ishihara2005}. 

The classical theory of turbulence predicts a mean scale at which the turbulent 
cascade ends and the flow viscosity starts to dominate. This scale is known as 
the Kolmogorov length \cite{Kolmogorov1941} 
\begin{equation}
\eta_K=\frac{\nu^{3/4}}{\langle\epsilon\rangle^{1/4}}\,,
\end{equation} 
where $\nu$ is the kinematic viscosity of the fluid. It is derived by a dimensional 
estimate and does not capture for the observed intermittency of the velocity gradients. 
We see that the energy dissipation rate, 
that probes the velocity gradient magnitude, enters the definition of 
$\eta_K$ as a mean,  $\langle\epsilon\rangle$. On the one hand, there are very 
intense gradients in the form of thin and stretched vortex tubes that seem to
have diameters around $\eta_K$, or even less.
On the other hand, ambient regions will exist with typical 
spatial variations larger than $\eta_K$. A whole range of local dissipation scales 
around the classical Kolmogorov length follows from the picture. This idea was put 
forward first within the multifractal formalism \cite{Paladin1987,Nelkin1990}. 
However, a direct analysis of the scales from data records remained very difficult, 
simply because these structures were not resolved. 

Yakhot derived an estimate that connects a dissipation scale $\eta$ with a 
velocity increment $u_{\eta}=|u(x+\eta)-u(x)|$ across this scale 
from the equations of motion (see \cite{Yakhot2006} and references therein). It reads
\begin{equation}
\eta u_{\eta}\approx \nu\,.
\label{eta}
\end{equation}
Technically, 
eq.~(\ref{eta}) is derived from a local kinetic energy balance by 
a so-called point-splitting procedure \cite{Polyakov1995,Lassig2000,Sreenivasan2005}.
Equation (\ref{eta}) tells us that $\eta$ is a field fluctuating in space and time 
and suggests an implicit way to determine the local dissipation scales $\eta$ in 
numerical simulations, by velocity increments over Kolmogorov and sub-Kolmogorov 
distances. Relation (\ref{eta}) can be also obtained by equating convective time scale, 
$\eta/u_{\eta}$, and dissipation time scale, $\eta^2/\nu$, a step
which is at the core of the multifractal approach of Paladin and Vulpiani \cite{Paladin1987}.  

\begin{figure}
\includegraphics[scale=0.25]{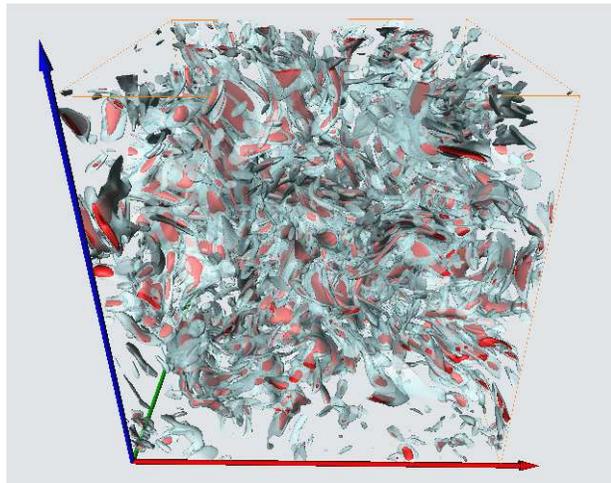}
\caption{(color online) Spatial distribution of local dissipation scales. An
instantaneous snapshot of a simulation with $1024^3$ grid points is shown (see Run 4
in table 1). Isosurfaces are plotted for two levels, $\eta=4\eta_K/3$ (shaded red) and
$5\eta_K/3$ (transparent gray). The isosurfaces for the smaller $\eta$ isolevel are nested
within those for the larger one and are significantly sparser distributed in space.
The figure underlines the fluctuating character of the local dissipation scale field.}
\label{fig.1}
\end{figure}

The probability density function (PDF) of the local 
dissipation scales will be denoted as $Q(\eta)$ in the following.
In fig.~\ref{fig.1}, we show an instantaneous snapshot of two isolevel sets of the field
$\eta$ where the isosurfaces in red are nested in the transparent gray ones. The figure 
clearly underlines the fluctuating character of the field $\eta$. 

Once one accepts the concept
of local dissipation scales, the question on the smallest local dissipation scale arises.
This scale can be determined by matching the inertial and dissipation range dynamics in the
equations for the longitudinal increment moments of order $2n$ which are defined as 
\begin{equation}
S_{2n}(r)=\langle u_r^{2n}\rangle = \langle|u(x+r)-u(x)|^{2n}\rangle
\label{increment}
\end{equation}
where $u$ is the turbulent velocity field projected onto ${\bf r}$ and 
$r=|{\bf r}|$ \cite{Sreenivasan2005,Schumacher2007}. The matching scale becomes order-dependent
and is given by
\begin{equation}
\eta_{2n}=L R^{\frac{1}{\zeta_{2n}-\zeta_{2n+1}-1}}.
\label{eta2n}
\end{equation}
Here, $L$ is the integral scale and $R=\sigma_L L/\nu$ the corresponding large-scale Reynolds
number with $\sigma_L=\langle u_L^2\rangle^{1/2}$. The exponents $\zeta_n$ determine
the scaling behaviour in the inertial range of turbulence, i.e. $S_n(r)\sim r^{\zeta_n}$ for
$\eta_K<r<L$. In the limit of $n\to\infty$ we expect $\zeta_{2n}\approx\zeta_{2n+1}$ and thus
\begin{equation}
\eta_{min}=L R^{-1}\,.
\label{etamin}
\end{equation}
One obtains a steeper decrease with respect to the Reynolds number
as for the Kolmogorov scale, $\eta_K=L R^{-3/4}$. The same relation (\ref{etamin}) follows 
in the multifractal approach by Paladin and Vulpiani \cite{Paladin1987} for the roughest 
increments with H\"older exponents $h=0$.       

\section{Numerical model}
The three-dimensional Navier-Stokes equations for an incompressible flow are solved.
The direct numerical simulations are based on the pseudospectral method with fast Fourier
transformations and a 2/3 de-aliasing.  The simulation domain is a periodic cube with a volume 
of $V=(2\pi)^3$. The velocity field is kept in a statistically stationary state by a large 
scale volume forcing that is added to the r.h.s. of the Navier-Stokes equations. It injects 
kinetic energy into the flow at a fixed rate, $\epsilon_{in}$, and thus prescribes the mean
energy dissipation rate. In the case of statistical stationarity the kinetic energy balance
reduces to $\epsilon_{in}=\langle\epsilon\rangle$. More details are found in \cite{Schumacher2007}. 
The turbulence is homogeneous and locally isotropic. Some parameters of the present
simulations are listed in table 1. The standard spectral resolution is determined by the criterion
$k_{max}\eta_K\ge 1.5$ \cite{Pope2000}. 
It is seen from the table that our applied resolution is at least by a 
factor of eight better than the standard case.

\renewcommand{\arraystretch}{1.4}
\begin{table}
\caption{Parameters of the direct numerical simulations 
$R_{\lambda}=\sqrt{15/(\langle\epsilon\rangle\nu)} \sigma_L^2$ 
is the Taylor-microscale Reynolds number, and $R=\sigma_L L/\nu$ 
the large scale Reynolds number with the integral scale $L$ and 
$\sigma_L^2=\langle (u_L)^2\rangle$ (see eq.~(\ref{increment})). The spectral 
resolution is indicated by $k_{max}\eta_K$ where $k_{max}=\sqrt{2}N/3 $.
The mean energy dissipation rate $\langle\epsilon\rangle$ is 0.1 for all
cases.}
\label{tab1}
\begin{center}
\begin{tabular}{lcccccc}
\hline
Run & $N$ & $\nu$ & $L$ & $R_{\lambda}$ & $R$ &  $k_{max}\eta_K$ \\
\hline
1   & 512  & 1/30   &  1.02   & 10   &   12    & 33.6   \\
2   & 1024 & 1/75   &  0.92   & 24   &   32    & 33.6   \\
3   & 1024 & 1/200  &  0.76   & 42   &   74    & 15.9   \\
4   & 1024 & 1/400  &  0.69   & 65   &  143    &  9.6   \\
5   & 2048 & 1/400  &  0.69   & 64   &  140    & 19.2   \\
6   & 2048 & 1/1000 &  0.66   & 107  &  347    &  9.7   \\
\hline
\end{tabular}
\end{center}
\end{table}

\section{Distribution of local dissipation scales}
The derivation of $Q(\eta)$ starts with relation (\ref{eta}) which connects 
velocities and scales at the small-scale end of the inertial range 
\cite{Yakhot2006}. $Q(\eta)$ can then be calculated from the PDF 
of the longitudinal velocity increments across distances $\eta$ {\em which is 
conditioned on} $u_{\eta}\eta/\nu=c$ (see eq.~(\ref{eta})). $c$ is a constant 
${\cal O}(1)$ and it follows 
\begin{equation}
Q(\eta)=\left|\frac{\mbox{d}u_{\eta}}{\mbox{d}\eta}\right|
P\left(u_{\eta}|\frac{u_{\eta}\eta}{\nu}=c\right)=
\frac{c\nu}{\eta^2} P\left(u_{\eta}|\frac{u_{\eta}\eta}{\nu}=c\right)\,.
\end{equation}
The PDF of the velocity increments is obtained 
from a Mellin transform \cite{Courant1989}. In Ref. \cite{Tcheou1999}, Mellin transforms were used
for the first time in turbulence to construct PDFs from increment moments by 
\begin{equation}
P(u_{\eta})=\frac{1}{i\pi u_{\eta}}\int_{-i\infty}^{+i\infty}\,\mbox{d}n\,
\,u^{-n}_{\eta} \,\langle u_{\eta}^n\rangle\,.
\label{mellin}
\end{equation}
Three 
points are necessary in order to make progress. Firstly, prefactors in the 
scaling laws for the increment moments are fixed by recognizing their Gaussian 
statistics at the largest scale of the flow, $L$. Secondly, the inertial cascade 
range scaling law for the increment moments is still valid at the small-scale end of the inertial 
range such that we can write
\begin{equation}
\langle u_{\eta}^{2p}\rangle=(2p-1)!! \,\sigma_L^{2p} \left(\frac{\eta}{L}\right)^{\zeta_{2p}}\,. 
\end{equation}
Thirdly, the unknown anomalous scaling exponents $\zeta_{2p}$ are approximated well 
with the polynomial 
$\zeta_{2p}=2ap-4bp^2$
for the lowest orders, $p<10$ \cite{Yakhot2006,Sreenivasan2005}. One finds 
$a=(1+9b)/3=0.383$ and $b=0.0166$ in order to satisfy the exact relation $\zeta_3=1$. 
The Mellin transformation integral is then evaluated by a saddle point approximation 
and results for a given large-scale Reynolds number $R$ to the following indefinite integral 
\begin{equation}
Q(\eta)=\frac{1}{\pi\eta\sqrt{b\log(L/\eta)}}\int_{-\infty}^{+\infty}\,\mbox{d}x
\exp\left[-x^2-\frac{\left(\log\left(\frac{\sqrt{2}x R}{c}\left(\frac{\eta}{L}\right)^{a+1}
\right)\right)^2}{4b\log(L/\eta)}\right]
\label{qeta}
\end{equation}
which can be analyzed by numerical quadrature. The distribution is supported for 
scales $0<\eta<L$ only.

\section{Comparison with numerical results}
\begin{figure}
\includegraphics[scale=0.55]{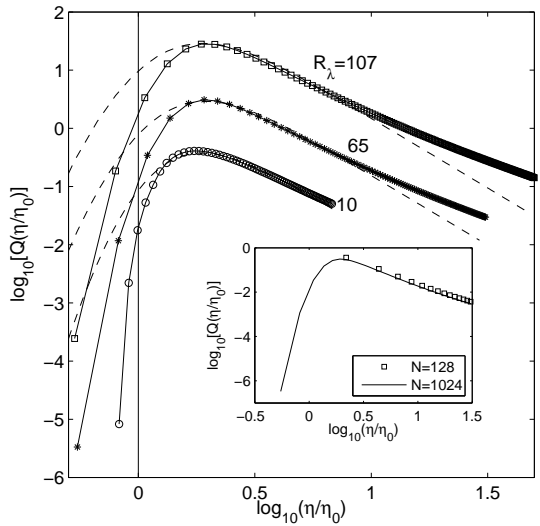}
\caption{Comparison of numerical and theoretical results from \cite{Yakhot2006} (dashed lines) 
for the 
probability density function of the local dissipation scale field $\eta$ 
at a given Reynolds number, $Q(\eta)$. Data are for Runs 1 (circles), 4 (asterisks) 
and 6 (squares) as given in Tab.~1. The values for $c$ are 2.6 (Run1), 4 (Run4), 
and 4 (Run6) respectively (see Eq.~(\ref{qeta})). All other parameters in DNS 
and (\ref{qeta}) are identical. For a better visibility, the data for Runs 4 and 
6 are shifted upwards by one and two orders of magnitude, respectively. 
Inset: Comparison 
of $Q(\eta)$ for the standard grid resolution with $k_{max} \eta_K=1.2$ and $N=128$ 
(squares) and the present very-high-resolution case (solid line). Data are for run 4.} 
\label{fig.2}
\end{figure}

The calculation of the PDF $Q(\eta)$ from the simulation data works as follows.
A scale $\ell$ is fixed as an integer multiple of the grid spacing $\Delta$, i.e.
$\ell=n \Delta$. The longitudinal velocity increments with respect to $\ell$ in 
all three directions are determined at each grid site. If 
$\ell u_{\ell}/\nu\approx  1$, the grid site contributes to $Q(\ell)$. The resulting 
distributions are shown in fig.~\ref{fig.2}. The inset underlines the necessity 
for the huge spectral resolution applied here. We compare a standard resolution 
case with the present one (which is eight times larger) while leaving all 
other parameters the same. The whole left tail of the local dissipation scale 
distribution cannot be resolved in the standard resolution case.

The main picture of the figure compares our data for Runs 1,4 and 6 (see table 1) with the 
theoretical prediction from (\ref{qeta}). The distributions are rescaled by
the scale $\eta_0$ that arises from eq. (\ref{eta2n}) when inserting $\zeta_{2p}=2ap-4bp^2$ to
\begin{equation}
\eta_{0}\simeq L R^{-\frac{1}{1+a}}=L R^{-0.72}\,,
\label{eta0}
\end{equation}
since $a=0.383$. We see that the distributions coincide 
quite well in the core and for most of the right tail of the PDF with the analytical shape. As stated 
in the caption, the parameter $c$ becomes independent of the Reynolds number $R$ which supports the
use of the scale $\eta_0$ for the rescaling. The smallest Reynolds
number case (Run 1) however deviates. The reason might be the Gaussian statistics of the
velocity gradients which changes to non-Gaussian for $R_{\lambda} > 10-15$
(see Ref.~\cite{Schumacher2007} for a detailed investigation on this subject). 

Furthermore, we observe that the distributions start to deviate at the largest scales 
from the analytical shape for the two larger Reynolds numbers. This is attributed to the 
periodic boundary conditions in the simulations which affect the velocity increments taken
over large distances. 
We have checked this by going up to velocity increments across the whole box length.  
Nevertheless, the slope of the algebraic decay for scales $\eta > \eta_K$ does not vary significantly 
with the Reynolds number, neither for the data nor for the theory. This part of the PDF 
which corresponds with increments over larger distances 
remains almost insensitive to an increase of Reynolds number. Note that all data collapse 
there. They have been shifted in the figure for a better visibility. Stronger deviations arise 
in the left tail, i.e. for the finest scales.
The analytical prediction (\ref{qeta}) is limited here. 
A next step would be to include higher-order corrections to the saddle-point approximation 
\cite{Tcheou1999}. 

\begin{figure}
\includegraphics[scale=0.55]{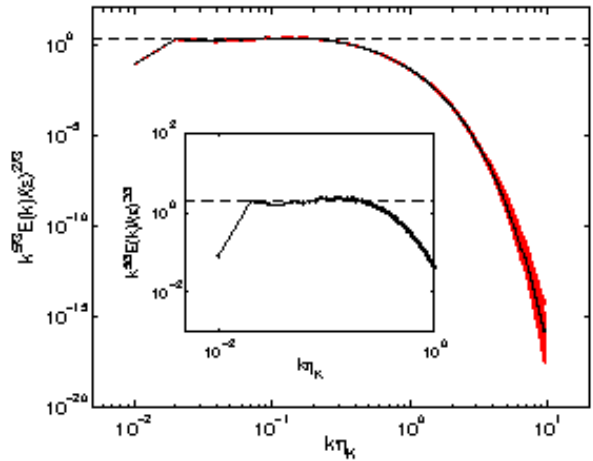}
\caption{(color online) Compensated energy spectrum for Run 6. The red lines are 
183 individual energy spectra saved during the time integration which took 4 large
scale eddy turnover times, $T_{eddy}=\langle u_i^2\rangle/L$. The black line is the 
mean of all these spectra which is used for the subsequent analysis. The dashed line 
is drawn at 2. The inset magnifies the range of wavenumbers that should show a 
bottleneck effect.} 
\label{fig.3}
\end{figure}
\begin{figure}
\includegraphics[scale=0.55]{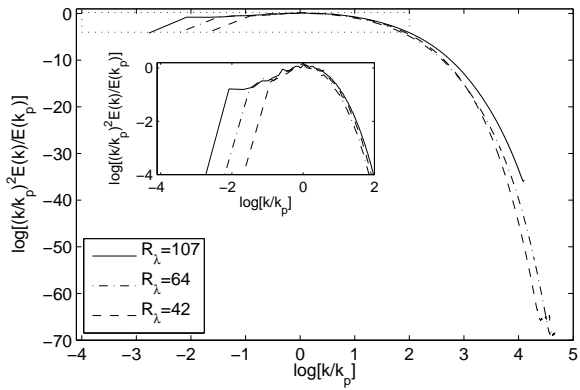}
\caption{Rescaled dissipation spectra for different Reynolds 
numbers. Wavenumbers are normalized by $k_p$ for which $k^2 E(k)$ becomes maximal. 
The dotted line box indicates the data window from \cite{Jackson1995} which is 
shown in the inset. The ratios $k_p\eta_K$ were found to 0.167, 0.179, and 0.160 
for $R_{\lambda}=42, 64$, and 107, respectively. The numerical precision of the DNS at 
such small amplitudes was tested by a verification of the viscous decay law of 
single Fourier modes without nonlinear advection, $\dot{\bf u}_{\bf k}(t)=-\nu 
k^2 {\bf u}_{\bf k}(t)$.} 
\label{fig.4}
\end{figure}
\begin{table}
\caption{Ratio of the finest local dissipation scale $\eta_{min}$ to the box size $L_x=2\pi$ following
from the DNS data and from the estimate in \cite{Yakhot2006}, respectively. $L_x$ is taken in order 
to show that $\eta_{min}^{DNS}/L_x$ is resolved, i.e. larger than $1/N_x$. Furthermore, it provides an independent scale being the same 
in all runs.}
\label{tab2}
\begin{center}
\begin{tabular}{lccccc}
\hline
$R_{\lambda}$ & 10 & 24 & 42 & 64 & 107 \\
\hline
$\eta_{min}^{DNS}$                     & $\frac{1}{51}$ & $\frac{1}{131}$ & $\frac{1}{407}$ &  $\frac{1}{576}$ & $\frac{1}{1194}$  \\
$\eta_{min}=L R^{-1}$ & $\frac{1}{74}$ & $\frac{1}{218}$ & $\frac{1}{613}$ & $\frac{1}{1295}$ & $\frac{1}{3232}$  \\
\hline
\end{tabular}
\end{center}
\end{table}
The numerical data for the left tail show a slight Reynolds number dependence. 
The value $Q(\eta/\eta_0=0.6)$ grows from approximately $10^{-7}$ at $R_{\lambda}=65$
to $10^{-5}$ at $R_{\lambda}=107$. The scales for the smallest Reynolds number go barely below 
$\eta_0$. This indicates an increasing probability of very fine sub-Kolmogorov scales 
to appear. It is in line with increasing small-scale intermittency of the velocity 
gradients which has been studied in \cite{Schumacher2007}. The small-scale end of the 
support in each of the PDFs is taken as the scale $\eta_{min}^{DNS}$. As discussed in the introduction, 
the theoretical model \cite{Yakhot2006,Schumacher2007} predicts  
$\eta_{min}= L R^{-1}$ (see eq.~(\ref{etamin})). Table 2 compares this estimate 
with our findings. We see that $\eta_{min}^{DNS}$ decreases more slowly than the 
theoretical prediction. The value of $\eta_{min}$ reached $0.6\eta_K$ for the largest 
Reynolds numbers. To conclude, ever finer sub-Kolmogorov scales are excited for increasing 
$R$, but not as pronounced as predicted by theory. 

The first moment of the distributions $Q(\eta)$ gives a mean dissipation scale which is
always larger than $\eta_K$. We find $\langle\eta\rangle=2.5,\, 6.2$ and 8.1 for $R_{\lambda}=10,
65$ and 107, respectively. For the range of accessible Reynolds numbers we thus observe an 
increase of the mean towards a scale $\sim 10\eta_K$ which is the generally expected 
crossover scale from the inertial to the viscous subrange \cite{Effinger1987}.

\section{Far-dissipation range energy spectra}
The central question that was addressed 
in \cite{Gagne1991,Sirovich1994,Lohse1995,Jackson1995,Frisch1991} is if the 
increasing small-scale intermittency in physical space manifests as well for 
the spectra in the crossover to the dissipation range or even in the far-dissipation 
range. Figure~\ref{fig.3} shows the instantaneous compensated energy spectra for the 
run at the largest Reynolds number. It provides information about the variation of the 
spectra in the dissipation range and how this adds up to a mean spectrum that will be 
used for the following analysis. 
Many numerical studies of the energy spectra have been focussed on the bottleneck 
phenomenon (see e.g. \cite{Minimi2006}). It is thought to result from a depletion 
of the nonlinear Fourier mode interaction at higher wavenumbers such that non-local
mode couplings become more dominant. We did not observe a bottleneck for our spectra (see
inset of fig.~\ref{fig.3}).
The reason can be the moderate Reynolds numbers which are accessed here. It can however
also be that the significantly larger range of resolved Fourier modes 
in the viscous range diminishes this effect.

Since the moderate Reynolds numbers prevent the quasi-algebraic scaling analysis 
from \cite{Frisch1991}, we study the spectral decay with respect to measures that 
follow from the spectra themselves \cite{Jackson1995}. The beginning of the dissipation 
range can be set at $k=k_p$ where $k^2E(k)$ has a maximum. In fig.~\ref{fig.4}, 
the dissipation spectra for the three largest $R$ are rescaled by the corresponding 
$k_p$. The dotted line indicates the data window from \cite{Jackson1995}. While 
the energy spectra collapse quite well within the dotted box, differences arise for 
the largest resolved wavenumbers. This indicates a deviation from the universality postulate made 
in the classical theory for the large-wavenumber decay of the energy spectra 
\cite{Kolmogorov1941}. We wish to stress that our Reynolds number might still be too 
small for a firmer conclusion.
     
\begin{figure}
\includegraphics[scale=0.55]{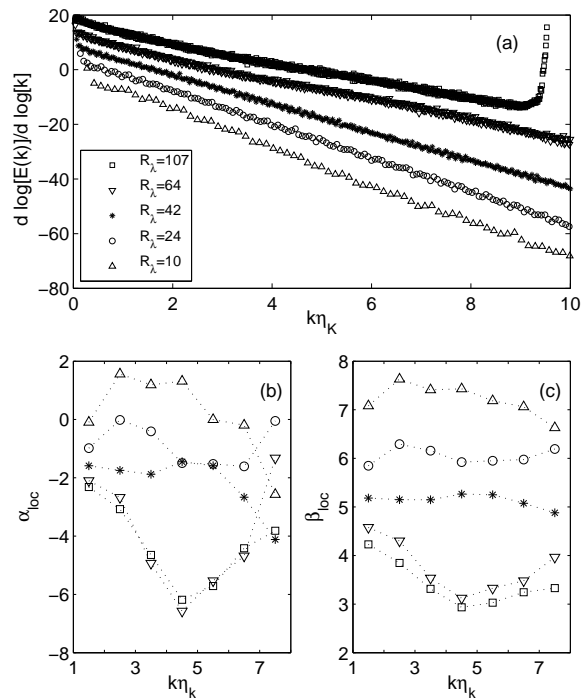}
\caption{Decay of the energy spectra $E(k)$ in the dissipation range for different
Reynolds numbers. (a) Local slope of the spectrum as a function of the wavenumber
$\tilde{k}=k\eta_K$. The symbols for the different Taylor microscale Reynolds numbers
are indicated in the legend of the figure. The spectra have been shifted  
with respect to each other for better visibility. (b) Local exponent 
$\alpha_{loc}$ as a function of the wavenumber $\tilde{k}$. Least square fits to (\ref{fit}) 
are done in the interval $[\tilde{k}-1/2, \tilde{k}+1/2]$. (c) Local exponent $\beta_{loc}$
as a function of $\tilde{k}$.} 
\label{fig.5}
\end{figure}
The results from fig.~\ref{fig.4} suggest a closer analysis of the Reynolds number dependence 
of the decay of the spectra in the far-dissipation range. The following form of the 
spectral decay for $\tilde{k}=k\eta_K> 1$ is used
\cite{Sirovich1994,Lohse1995,Chen1993,Martinez1997,Ishihara2005}
\begin{equation}
\tilde{E}(k)=\frac{E(k)}{\nu^{5/4}\langle\epsilon\rangle^{1/4}}=F(\tilde{k})=
\tilde{k}^{\alpha}\exp(-\beta \tilde{k})
\label{decay}
\end{equation}
where $\alpha$ and $\beta$ are Reynolds-number-dependent dimensionless constants.
The exponential term can be motivated from Stokes eigenfunctions in the 
viscosity-dominated regime of the Navier-Stokes dynamics \cite{Foias1990}.
The power law term $\tilde{k}^{\alpha}$ was calculated by the direct interaction
approximation (DIA) to $\alpha=3$ \cite{Kraichnan1959}. Numerical experiments at low Reynolds
numbers demonstrated later that $\alpha$ can exceed 3 due to small-scale intermittency
which is not contained in the theory \cite{Chen1993}.    
Equation (\ref{decay}) is transformed into the following local slope form  
\begin{equation}
\frac{\mbox{d}\log(\tilde{E}(k))}{\mbox{d}\log(\tilde{k})}=\alpha-\beta\tilde{k}
\label{fit}
\end{equation}
which allows for direct determination of the two constants $\alpha$ and $\beta$ by a 
least square fit. The upper picture of fig.~\ref{fig.5} shows the local slope of the
large-wavenumber spectral decay. The two lower pictures in the same figure list the results 
for both coefficients, $\alpha_{loc}$ and $\beta_{loc}$. We have performed therefore {\em local} least
square fits to (\ref{fit}) in the interval $[\tilde{k}-1/2,\tilde{k}+1/2]$.   
It should be stressed once more that the present data allow a systematic study of the dissipation 
range decay over an order of magnitude of Taylor microscale Reynolds numbers. The results
are consistent with the findings from Refs. \cite{Chen1993,Martinez1997,Ishihara2005}.
The exponent $\alpha$, which determines the nature of the singularity in the Euler case,
remains smaller zero for $R_{\lambda}>24$ which results in a convex shape of the dissipation range
spectrum. Interestingly, this is the range of $R_{\lambda}$ for which velocity gradients
are clearly non-Gaussian \cite{Schumacher2007}.
The overall magnitude of the exponent $\beta$ decreases for growing Reynolds number. 
A saturation of this decrease can be detected for the two largest Reynolds number runs. 
This would be consistent with a saturation to a constant magnitude which was predicted by Kraichnan 
\cite{Kraichnan1959}. 
Neither for $\alpha$ nor for $\beta$ a systematic behaviour with 
respect to wavenumber $\tilde{k}$ is observed. The limited range of sub-Kolmogorov
scales that can be resolved is one reason. Moreover, it should be stressed  
that the local fits of the exponential decay require a very high numerical accuracy as has been 
demonstrated in Pauls {\it et al.} \cite{Pauls2006}. The double precision floating accuracy 
which we can provide only was not sufficient to apply more sophisticated local fit procedures 
to (\ref{decay}) \cite{Pauls2006,Shelley1993}. Finally, it was also checked that the rescaling 
with $\eta_0$ instead of $\eta_K$ does not alter the fit results significantly. 

\section{Concluding remarks} 
The finest-scale intermittent fluctuations of fluid turbulence have been studied with a 
spectral resolution never applied before. 
They are associated with a whole range of local dissipation scales rather than a
mean dissipation scale -- the Kolmogorov length $\eta_K$. We find that the increase of small-scale
intermittency with increasing Reynolds number goes in line with an increasing spread-out 
of the local dissipation scales into the dissipation range. The growth of this scale range
with respect to the Reynolds number is small, but present. 
This is consisitent with a logarithmic dependence
of the extension of the intermediate dissipation range on the Reynolds number as proposed in
\cite{Chevillard2005}.   Furthermore, we detected a growing amplitude 
of sub-Kolmogorov-scale fluctuations which manifests in a slower exponential decay of the energy 
spectra in the large-wavenumber range. Both numerical results confirm indications from other
studies that increasing  
small-scale intermittency affects a growing number of scales in the dissipation range that are expected
to be Reynolds-number-independent (and thus universal) in the classical theory of turbulence. 

\acknowledgments
The supercomputing time was provided by the Deep Computing 
Initiative of the Distributed European Infrastructure for Supercomputer Applications 
consortium (DEISA) on the JUMP cluster at the John von Neumann Institute for Computing, 
J\"ulich (Germany). The author thanks for this support and is grateful to L. Biferale,
U. Frisch, T. Ishihara, Y. Kaneda, K. R. Sreenivasan, A. Thess and V. Yakhot for comments and suggestions. This work was also supported by the Deutsche Forschungsgemeinschaft with Grant No. 
SCHU 1410/2 and the German Academic Exchange Service (DAAD).

\end{document}